\documentstyle[twoside,fleqn,espcrc2]{article}

\newcommand{\AmS}{{\protect\the\textfont2
  A\kern-.1667em\lower.5ex\hbox{M}\kern-.125emS}}

\newcommand{\be}{\begin{equation}}
\newcommand{\ee}{\end{equation}}
\newcommand{\ba}{\begin{eqnarray}}
\newcommand{\ea}{\end{eqnarray}}

\newcommand{\Tr}{{\rm Tr}}
\newcommand{\Dmrns}{{\cal D}_{\mu\rho,\nu\sigma}}

\title{Field strength correlators in QCD with dynamical fermions\thanks{
Partially supported by MURST and by the EC contract CHEX--CT92--0051.}}

\author{M. D'Elia\address{Dipartimento di Fisica and INFN,
        Piazza Torricelli 2, 56100 Pisa, Italy},
        A. Di Giacomo$^{\rm a}$,
        E. Meggiolaro$^{\rm a}$\thanks{Speaker at the conference.}
       }
       
\begin{document}

\begin{abstract}
We determine, by numerical simulations on a lattice, the gauge--invariant 
two--point correlation functions of the gauge field strengths in the QCD 
vacuum with four flavours of dynamical staggered fermions.
\end{abstract}

\maketitle

The gauge--invariant two--point correlators of the gauge field strengths in 
the QCD vacuum are defined as
\be
\Dmrns(x) = \langle 0| 
\Tr \left\{ G_{\mu\rho}(x) S G_{\nu\sigma}(0) S^\dagger \right\}
|0\rangle ~,
\ee
where $G_{\mu\rho} = gT^aG^a_{\mu\rho}$ is the field--strength tensor and
$S = S(x,0)$ is the Schwinger phase operator needed to 
parallel--transport the tensor $G_{\nu\sigma}(0)$ to the point $x$.

These correlators govern the effect of the gluon condensate on the level 
splittings in the spectrum of heavy $Q \bar{Q}$ bound states
\cite{Gromes82,Campostrini86,Simonov95}.
They are the basic quantities in models of stochastic confinement of colour
\cite{Dosch87,Dosch88,Simonov89}
and in the description of high--energy hadron scattering
(see Ref. \cite{Dosch94} and references therein).

A numerical determination of the correlators on the lattice 
already exists in the {\it quenched} 
(i.e., pure--gauge) $SU(2)$ theory \cite{Campostrini84}, 
and also in the {\it quenched} $SU(3)$ theory \cite{DiGiacomo92,myref97}.
Here we present results obtained in {\it full} QCD, i.e., by 
including the effects of dynamical fermions \cite{newref97}.
Four flavours of {\it staggered} fermions and the Wilson action for the 
pure--gauge sector have been used. The determination has been done at two 
different values of the quark mass.

In the Euclidean theory, the most general parametrization of the correlators 
is the following \cite{Dosch87,Dosch88,Simonov89}:
\ba
\lefteqn{
\Dmrns(x) = (\delta_{\mu\nu}\delta_{\rho\sigma} - 
\delta_{\mu\sigma}\delta_{\rho\nu})
\left[ {\cal D}(x^2) + {\cal D}_1(x^2) \right] } \nonumber \\
& & + (x_\mu x_\nu \delta_{\rho\sigma} - x_\mu x_\sigma \delta_{\rho\nu} 
+ x_\rho x_\sigma \delta_{\mu\nu} - \nonumber \\
& & x_\rho x_\nu \delta_{\mu\sigma})
{\partial{\cal D}_1(x^2) \over \partial x^2} ~,
\ea
where ${\cal D}$ and ${\cal D}_1$ are invariant functions of $x^2$.
It is also convenient to define the following quantities: 
\ba
\lefteqn{
{\cal D}_\parallel \equiv {\cal D} + {\cal D}_1 + x^2 {\partial{\cal D}_1
\over \partial x^2} ~,} \nonumber \\
\lefteqn{
{\cal D}_\perp \equiv {\cal D} + {\cal D}_1 ~.}
\ea
On the lattice we can define an operator $\Dmrns^L$, which is
proportional to $\Dmrns$ in the na\"\i ve continuum limit, i.e., when the 
lattice spacing $a \to 0$ \cite{DiGiacomo92,myref97}:
\ba
\lefteqn{
{\cal D}_{\parallel,\perp}^L(\hat d a) \mathop\sim_{a\to0} a^4 
{\cal D}_{\parallel,\perp} (d^2 a^2) + {\cal O}(a^6) ~.} 
\label{naive}
\ea
However, the na\"\i ve continuum limit of Eq. (\ref{naive}) is spoiled by 
the presence of lattice artefacts, i.e., renormalization effects from 
lattice to continuum due to the short--range fluctuations at the scale of the 
UV cutoff. In order to remove these artefacts we adopt the same technique
used in Refs. \cite{DiGiacomo92,myref97}. 
The basic idea  is to remove the effects of short--range fluctuations on 
large distance correlators by a local {\it cooling} procedure
\cite{Campostrini89,DiGiacomo90}.

We have measured the correlations on a $16^3 \times 24$ lattice at distances 
$d$ ranging from 3 to 8 lattice spacings and at $\beta = 5.35$ 
($\beta = 6/g^2$, where $g$ is the coupling constant).
We have used a standard Hybrid Monte Carlo algorithm,
in particular the so--called $\Phi$--algorithm described in Ref.
\cite{Gottlieb87}.
The bare quark mass was chosen to be $a \cdot m_q = 0.01$. 
A determination was also made at $a \cdot m_q = 0.02$. 

The scale of our system is fixed by the physical value of the lattice spacing 
$a$. We shall use the following  parametrization
\be
a(\beta) = {1\over\Lambda_F} \left({8\over25}\,\pi^2\beta\right)
^{ 231/625 } \exp\left(-{4\over25}\pi^2\beta\right) ,
\label{scale}
\ee
where the {\it scaling function} $f(\beta) = \Lambda_F \cdot a(\beta)$ is given 
by the usual two--loop expression for gauge group $SU(3)$ and $N_f = 4$ 
flavours of quarks.
$\Lambda_F$ in Eq. (\ref{scale}) is an effective $\Lambda$--parameter for QCD 
in the lattice renormalization scheme, with $N_f = 4$ flavours of quarks.
At the value of $\beta$ used in our simulation the lattice spacing, extracted 
from the string tension or the $\rho$ mass, is $a \simeq 0.11 \pm 0.01$ fm  
\cite{Laermann93,Laermann94}, so that our lattice size is approximately 2 fm 
and therefore safe from infrared artefacts.

\begin{figure}[t]
\vspace{5.0cm}
\includegraphics{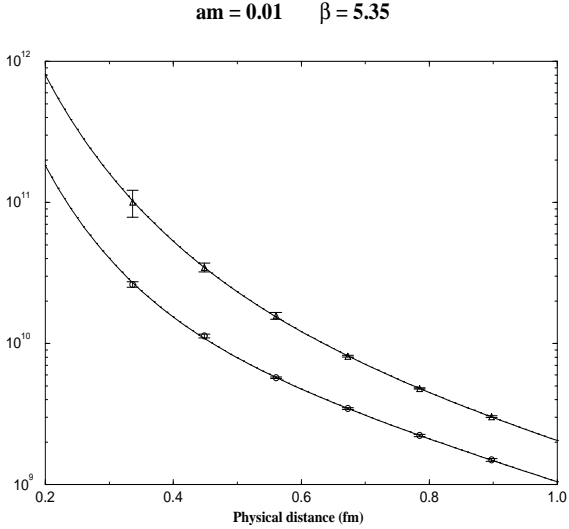} 
\null\vskip 0.7cm
\caption{The functions ${\cal D}_\perp /\Lambda_F^4$ 
(upper curve) and ${\cal D}_\parallel /\Lambda_F^4$ (lower curve)
versus physical distance, for quark mass $a \cdot m_q = 0.01$.}
\end{figure}

\begin{figure}[t]
\vspace{5.0cm}
\includegraphics{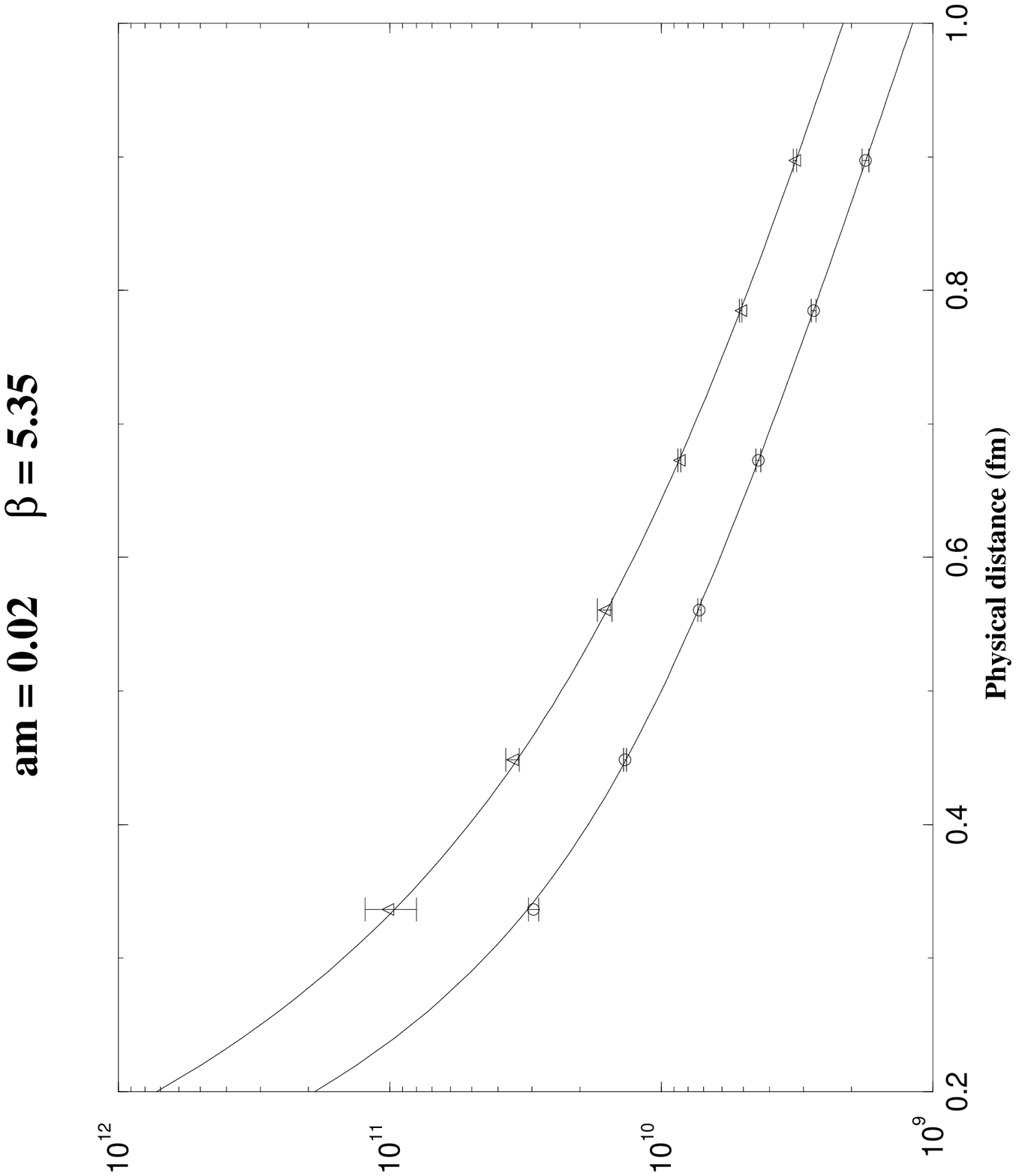} 
\null\vskip 0.7cm
\caption{The functions ${\cal D}_\perp /\Lambda_F^4$ 
(upper curve) and ${\cal D}_\parallel /\Lambda_F^4$ (lower curve)
versus physical distance, for quark mass $a \cdot m_q = 0.02$.}
\end{figure}

In Fig. 1 we display the results for ${\cal D}_\parallel /\Lambda_F^4$ 
and ${\cal D}_\perp /\Lambda_F^4$ versus $d_{\rm phys} = d \cdot a$,
for a quark mass $a \cdot m_q = 0.01$.
As in Ref. \cite{myref97}, we have tried a best fit to these data with the 
functions
\ba
\lefteqn{
{\cal D} (x^2) = A_0 \exp\left( -{{|x|}\over{\lambda_A}} \right) 
+ {a_0 \over |x|^4} \exp\left( -{{|x|}\over{\lambda_a}} \right) ~,}
\nonumber \\
\lefteqn{
{\cal D}_1 (x^2) = A_1 \exp\left( -{{|x|}\over{\lambda_A}} \right)
+ {a_1 \over |x|^4} \exp\left( -{{|x|}\over{\lambda_a}} \right) .}
\label{parmalat}
\ea
The continuum lines in Fig. 1 correspond to the central values of this best 
fit (see Ref. \cite{newref97}).

The corresponding results for the quark mass $a \cdot m_q = 0.02$ are 
displayed in Fig. 2. From $a \cdot m_q = 0.01$ to $a \cdot m_q = 0.02$ the 
effective $\Lambda_F$ does not change appreciably within the 
errors, so that we have assumed the same value of $\Lambda_F$ as for 
$a \cdot m_q = 0.01$.
Again, the continuum lines in Fig. 2 correspond to the central values of the 
best fit with the functions (\ref{parmalat}) (see Ref. \cite{newref97}).

A quantity of physical interest which can be extracted from our lattice 
determination is the correlation length $\lambda_A$ of the gluon field 
strengths, defined in Eq. (\ref{parmalat}): it is relevant for the description 
of vacuum models \cite{Dosch87,Dosch88,Simonov89}.
To obtain the value of $\lambda_A$ in physical units, the physical value of the
lattice spacing must be used. This gives, for the first quark mass 
$a \cdot m_q = 0.01$:
\be
\lambda_A = 0.34 \pm 0.02 \pm 0.03 \, {\rm fm} ~~~~~~ (a \cdot m_q = 0.01) 
~,
\label{lamb1}
\ee
where the first error comes from our determination and the second 
from the uncertainty in fixing the physical scale.

Similarly, for the second quark mass $a \cdot m_q = 0.02$ we obtain:
\be
\lambda_A = 0.29 \pm 0.01 \pm 0.03 \, {\rm fm} ~~~~~~ (a \cdot m_q = 0.02)
~.
\label{lamb2}
\ee
The values (\ref{lamb1}) and (\ref{lamb2}) must be compared with the 
{\it quenched} value \cite{DiGiacomo92,myref97}
\be
\lambda_A = 0.22 \pm 0.01 \pm 0.02 \, {\rm fm} ~~~~~~ ({\rm YM ~ theory}) ~.
\ee
The correlation length $\lambda_A$ decreases by increasing the quark mass,
when going from chiral to {\it quenched} QCD.
Of course, a determination of the physical value of $\lambda_A$ should be done 
with more realistic values  for the quark masses.

Another quantity of physical interest which can be extracted from our results 
is the  the so--called {\it gluon condensate}, defined as
\be
G_2 \equiv \langle {\alpha_s \over \pi} :G^a_{\mu\nu} G^a_{\mu\nu}: \rangle
~~~~~~~~ (\alpha_s = {g^2 \over 4\pi}) ~.
\ee
As first pointed out by Shifman, Vainshtein and Zakharov \cite{SVZ79}, 
it is a fundamental quantity in QCD, in the context of the sum rules.

The gluon condensate can be expressed, in terms of the parameters defined in 
Eq. (\ref{parmalat}), as follows \cite{newref97}:
\be
G_2 \simeq {6 \over \pi^2} (A_0 + A_1) ~.
\ee
At $a \cdot m_q = 0.01$ this gives, in physical units,
\be
G_2 = 0.015 \pm 0.003 \,^{+0.006}_{-0.003} \, {\rm GeV}^4 ~.
\label{first_val}
\ee
At $a \cdot m_q = 0.02$ we obtain:
\be
G_2 = 0.031 \pm 0.005 \,^{+0.012}_{-0.007} \, {\rm GeV}^4 ~.
\label{second_val}
\ee
These values should be compared with the corresponding {\it quenched} value 
\cite{myref97}:
\be
G_2 = 0.14 \pm 0.02 \,^{+0.06}_{-0.05}  \, {\rm GeV}^4 ~~~~~~ 
({\rm YM ~ theory}) ~.
\label{quench_val}
\ee
As expected, the gluon condensate $G_2$ appears to increase with the quark 
mass, tending towards the (pure--gauge) value of Eq. (\ref{quench_val}). 
We can try to understand the dependence of $G_2$ on the 
quark masses using  the following 
low--energy theorem \cite{NSVZ81}, valid for small quark masses:
\be
{d \over dm_f} \langle {\alpha_s \over \pi} :G^a_{\mu\nu}G^a_{\mu\nu}:
\rangle = -{24 \over b} \langle :\bar{q}_f q_f: \rangle ~,
\ee
where $b = 11 - {2 \over 3} N_f$, for a gauge group $SU(3)$ and $N_f$ quark 
flavours. 
For $a \cdot m_q = 0.01$ we have 
approximately $m_f \simeq 44$ MeV \cite{Laermann93}.
Making use of the popular values for the quark condensate ($\langle \bar{q} 
q \rangle \simeq -0.013 \, {\rm GeV}^3$ \cite{Laermann93,Dosch95}) and for the 
physical quark masses ($m_u \simeq 4$ MeV, $m_d \simeq 7$ MeV and $m_s \simeq 
150$ MeV), we can extrapolate from the value (\ref{first_val}) to the 
{\it physical} gluon condensate, obtaining the following estimate:
\be
G_2^{(physical)} \sim 0.022 \, {\rm GeV}^4 ~.
\label{phys_val}
\ee
The prediction (\ref{phys_val}) agrees with phenomenological  determinations 
\cite{Dosch95,Narison96}: 
$G_2^{(empiric)} \simeq 0.024 \pm 0.011 \, {\rm GeV}^4$.

\end{document}